\documentclass[aps,prl,twocolumn,floatfix,superscriptaddress,showpacs,
amsfonts,amssymb,amsmath,preprintnumbers,10pt]{revtex4-1}

\usepackage{bm}

\newcommand{\Exref}[1]{(\ref{#1})}
\newcommand{\Eqref}[1]{Eq.~(\ref{#1})}
\newcommand{\Eqsref}[1]{Eqs.~(\ref{#1})}

\newcommand{\Eqsand}[2]{Eqs.~(\ref{#1}) and (\ref{#2})}

\renewcommand{\vec}[1]{\bm{#1}}

\newcommand{\vdel}{\vec{\nabla}}
\newcommand{\unity}{\mathbf{\hat{I}}}

\newcommand{\vk}{\vec{k}}
\newcommand{\uzk}{u_{z\vk_0}}
\newcommand{\uzkk}{u_{z\vk_0'}}
\newcommand{\Phik}{\Phi_{\vk_0}}
\newcommand{\Phikk}{\Phi_{\vk_0'}}
\newcommand{\vBp}{\vec{B}^\perp}
\newcommand{\vBk}{\vBp_{\vk_0}}
\newcommand{\Bxk}{B_{x\vk_0}}
\newcommand{\vBo}{\vec{\bar{B}}}
\newcommand{\Bxo}{\bar{B}_x}
\newcommand{\Byo}{\bar{B}_y}
\newcommand{\vBmk}{\vBp_{-\vk_0}}
\newcommand{\vBkk}{\vBp_{\vk_0-\vk_0'}}
\newcommand{\Fk}{F_{\vk_0}}
\newcommand{\fzk}{f_{z\vk_0}}
\newcommand{\talpha}{\vec{\hat{\alpha}}}
\newcommand{\tA}{\bm{\hat{\vec{A}}}}
\newcommand{\tS}{\bm{\hat{\vec{S}}}}
\newcommand{\tD}{\bm{\hat{\vec{D}}}}
\newcommand{\vC}{\bm{\vec{C}}}

\newcommand{\ey}{\vec{e}_y}
\newcommand{\ez}{\vec{e}_z}

\newcommand{\lf}{\ell_f}
\newcommand{\tcorr}{\tau_c}
\newcommand{\kpeak}{k_z^\mathrm{pk}}

\newcommand{\Rey}{\mathrm{Re}}
\newcommand{\Rm}{\mathrm{Rm}}

\newcommand{\Sh}{\mathrm{Sh}}
\newcommand{\St}{\mathrm{St}}
\newcommand{\Ku}{\mathrm{Ku}}
\newcommand{\rmi}{\mathrm{i}}
\newcommand{\rme}{\mathrm{e}}
\newcommand{\rmd}{\mathrm{d}}
\newcommand{\urms}{u_\mathrm{rms}}

\begin{document}

\title{Large-Scale Magnetic Field Generation by Randomly Forced Shearing Waves}
\author{T.~Heinemann}
\affiliation{Institute for Advanced Study, Princeton, New Jersey 08540, USA}
\author{J.~C.~McWilliams}
\affiliation{Department of Atmospheric Sciences, UCLA, Los Angeles, California 90095-1565, USA}
\author{A.~A.~Schekochihin}
\affiliation{Rudolf Peierls Centre for Theoretical Physics, University of Oxford, Oxford OX1 3NP, United Kingdom}

\date{\today}

\begin{abstract}
A rigorous theory for the generation of a large-scale magnetic field by random
non-helically forced motions of a conducting fluid combined with a linear
shear is presented in the analytically tractable limit of low $\Rm$ and weak
shear. The dynamo is kinematic and due to fluctuations in the net
(volume-averaged) electromotive force. This is a minimal proof-of-concept
quasilinear calculation aiming to put the shear dynamo, a new effect recently
found in numerical experiments, on a firm theoretical footing. Numerically
observed scalings of the wavenumber and growth rate of the fastest growing
mode, previously not understood, are derived analytically. The simplicity of
the model suggests that shear dynamo action may be a generic property of
sheared magnetohydrodynamic turbulence.
\end{abstract}


\maketitle

\paragraph{Introduction.}
Magnetogenesis, or origin of cosmic magnetism, is one of the fundamental
problems in theoretical astrophysics. It has long been believed that the
magnetic fields observed in most astrophysical bodies owe their existence to
the dynamo effect associated with the turbulence of the constituent plasmas.
It is not controversial that turbulence of a conducting fluid amplifies
magnetic fluctuations at scales comparable to or smaller than the scale of the
motions. Small-scale magnetic fluctuations are indeed observed ubiquitously,
but in most astrophysical systems, one also finds magnetic fields coherent
on scales larger than the scale of the turbulence (e.g., \cite{Shukurov}).
Generation of such fields, or mean-field dynamo action, is expected to require
a combined action of turbulence and some large-scale-coherent feature. One
well-known such additional ingredient is net kinetic helicity (or, more
generally, reflectional asymmetry) of the motion. Under certain conditions,
its presence can cause growth of large-scale (``mean'') magnetic field, known
as the $\alpha$-effect \cite{Steenbeck_Krause_Raedler,*Moffatt_book}. While
deriving the $\alpha$-effect for realistic turbulent systems requires rather
drastic closure assumptions, which usually cannot be justified rigorously and
have, in fact, been called into question by numerical and analytical
considerations \cite{Boldyrev_Cattaneo_Rosner}, it is at least clear that the
effect exists in the physically realizable and analytically treatable limit of
low $\Rm$ \cite{Steenbeck_Krause_Raedler,*Moffatt_book,Raedler_Rheinhardt}.
This proof-of-concept analytical result, together with intuitive arguments
\cite{Parker} and a body of numerical evidence
\cite{Brandenburg_alpha,BrandenburgNordlund}, have helped build a case for the
$\alpha$-effect as a real physical phenomenon (although whether it can coexist
with the small-scale dynamo at large $\Rm$ is far from certain
\cite{Boldyrev_Cattaneo_Rosner}). 

It has been suggested
\cite{Vishniac_Brandenburg,Blackman,Urpin_mnras99,RK03,Brandenburg2005} that
even in the absence of \emph{mean} helicity, mean-field dynamo action is
possible if a large-scale velocity shear is present. The importance of such a
possibility can hardly be overestimated, as shear is a ubiquitous feature in
astrophysics (usually associated with differential rotation). A recent
numerical study \cite{Yousef_etal_PRL,Yousef_etal_AN} showed that the shear
dynamo does exist, but its nature has remained poorly understood. The
uncertainty is increased by the fact that, while the original derivation of
the effect relied on a quantitative outcome of a closure calculation
\cite{RK03}, the effect proved difficult to identify by numerical computation
of the mean-field-theory coefficients \cite{Brandenburg_etal_stoch_alpha} and
appeared to go away in rigorously solvable limits: the white-noise-velocity
model and low-$\Rm$ magnetohydrodynamics
\cite{Raedler_Rheinhardt,Ruediger_Kichatinov,Kichatinov_AN,
SridharSubramanian2009a,*SridharSubramanian2009b,
*SridharSingh2010,*SridharSingh2011} (but see \cite{RK08}). 

In this Letter, our aim is a minimal proof-of-concept calculation that puts
the shear dynamo effect on a firm theoretical footing akin to that enjoyed by
the $\alpha$-effect. We propose a very simple quasilinear mean-field theory
that rigorously predicts a large-scale dynamo driven by randomly forced
shearing waves in the limit of \mbox{$\Rm\ll\Rey\ll1$}. The effect requires no
adjustable parameters. We also recover the scalings of the wavenumber and
growth rate of the fastest-growing mode that were observed in a number of
numerical studies
\cite{Yousef_etal_PRL,Yousef_etal_AN,Kapyla_etal,*Hughes_Proctor} but have not
so far been explained analytically. 

\paragraph{Shearing Waves.}
First let us introduce a model velocity field that will be used to obtain a
dynamo. Consider an incompressible fluid with an imposed background linear
shear, $\vec{U}=Sx\ey$,  and assume that the magnetic field is dynamically
weak, so the Lorentz force is negligible. Then the velocity deviation from
$\vec{U}$ satisfies
\begin{equation}
\label{NSEq}
\partial_t\vec{u} + Sx\partial_y\vec{u} + S u_x\ey + \vec{u}\cdot\vdel\vec{u} = -\vdel p +
\nu\nabla^2\vec{u} + \vec{f},
\end{equation}
where $p$ is pressure determined from incompressibility
\mbox{$\vdel\cdot\vec{u}=0$}, $\nu$ is viscosity and $\vec{f}$ is a random
body force, assumed to be statistically homogeneous in time and space and to
have a characteristic scale $\lf$. 

We now make two simplifying assumptions. First, let
$\Rey\sim\urms\lf/\nu\ll1$, so  we can neglect the nonlinear term in
\Eqref{NSEq}. Second, let $\partial_z\vec{u}=0$ and $\partial_z\vec{f}=0$,
resulting in a ``quasi-2D'' velocity with all three vector components but no
$z$-dependence. This velocity will make our calculations particularly
transparent. As indicated by numerical experiments \cite{Mallet_Yousef}, it is
a favorable but not a uniquely special case as a dynamo. The $xy$-plane
velocity now has a stream function: $\vec{u}^\perp = \ez\times\vdel\Phi$.
Similarly, $\vec{f}^\perp=\ez\times\vdel F$. We seek solutions of (the
linearized) \Eqref{NSEq} as superpositions of ``shearing waves''
\cite{Kelvin_shwaves}:
\begin{equation}
\label{shwaves_decomp}
\Phi = \sum_{\vk_0}\!\Phik(t)\,\rme^{\rmi\vk(t)\cdot\vec{r}},
\quad
u_z = \sum_{\vk_0}\!\uzk(t)\,\rme^{\rmi\vk(t)\cdot\vec{r}},
\end{equation} 
where $\vk_0 = k_{x0}\vec{e}_x + k_y\vec{e}_y$, $\vk(t)=(k_{x0} - St
k_y)\vec{e}_x + k_y\vec{e}_y$. The amplitudes of the shearing waves satisfy
\footnote{ Note that when the body force $\vec{f}$ is decomposed into shearing
waves, the forcing amplitudes $\Fk(t)$ and $\fzk(t)$ are of significant
magnitude only when $k(t)\sim\lf^{-1}$, because we want energy injection to
occur at a scale $\lf$ in the laboratory frame $(x,y,t)$.}
\begin{align}
\label{Phi_eq}
\partial_t\left[k^2(t)\Phik\right] &= -\nu k^4(t)\Phik + k^2(t)\Fk,\\
\label{uz_eq}
\partial_t \uzk &= -\nu k^2(t)\uzk + \fzk.
\end{align}
\Eqref{Phi_eq} was obtained by taking
$\ez\cdot\left[\vdel\times{\rm\Eqref{NSEq}}\right]$. For simplicity, let us
consider the forcing to be white in time (or, equivalently, to have a
correlation time much shorter than the viscous relaxation time $\lf^2/\nu$).
Then the two-point velocity correlators are
\begin{align}
\label{Phi12_corr}
\bigl<\Phik(t)\Phikk^*(t')\bigr> &= \delta_{\vk_0,\vk_0'}G_\nu(t,t')
\frac{k^2(t')}{k^2(t)}
\bigl<|\Phik(t')|^2\bigr>,\\
\label{uz12_corr}
\!\!\bigl<\uzk(t)\uzkk^*(t')\bigr> &=
\delta_{\vk_0,\vk_0'}G_\nu(t,t')\bigl<|\uzk(t')|^2\bigr>,
\end{align}
where $G_\nu(t,t') = \exp\left[-\nu\int_{t'}^t\rmd t''k^2(t'')\right]$. Thus,
the correlation time of our velocity field is \mbox{$\tcorr\sim\lf^2/\nu$}. In
a more general case when $\Rey$ is not small, the velocity correlation time is
set by the nonlinear terms, so \mbox{$\tcorr\sim\lf/\urms$} is the typical
turnover time of the turbulence. Non-rigorously, this case is included in our
analysis. To accommodate it, we introduce the Strouhal number
$\St\sim\urms\tcorr/\lf$ (following \cite{Raedler_Rheinhardt}) --- then
$\St\sim\Rey$ for a velocity governed by \Eqsand{Phi_eq}{uz_eq}, and
$\St\sim1$ for conventional turbulence.

An important quantity to watch is the net (volume-, but not time-, averaged)
helicity 
\begin{equation}
  \label{helicity}
  \mathcal{H}(t) = \langle\vec{u}\cdot(\vdel\times\vec{u})\rangle_{xy} =
  -2\sum_{\vk_0}k^2(t)\uzk(t)\Phik^*(t).
\end{equation}
We can ensure that its statistical (or, equivalently, time) average vanishes,
$\langle\mathcal{H}(t)\rangle=0$, by stipulating
\mbox{$\left<\fzk(t)\Fk^*(t')\right>=0$}. This removes the possibility of the
standard $\alpha$-effect
\cite{Steenbeck_Krause_Raedler,*Moffatt_book,Raedler_Rheinhardt}.

\paragraph{Mean-Field Theory.}
The evolution equation for the magnetic field $\vec{B}$ in the presence of
linear shear is 
\begin{equation}
  \label{induction}
  \partial_t\vec{B} + Sx\partial_y\vec{B} + \vec{u}\cdot\vdel\vec{B} =
  \vec{B}\cdot\vdel\vec{u} + SB_x\ey + \eta\nabla^2\vec{B},
\end{equation}
where $\eta$ is the magnetic diffusivity. Since the velocity field is
independent of $z$, we can separate the dependence of $\vec{B}$ on $z$ by
expanding \mbox{$\vec{B}=\sum_{k_z}\vec{B}(k_z)\exp(\rmi k_z z)$}. Only the
projection $\vBp$ onto the $xy$-plane needs to be calculated because
\mbox{$B_z=(\rmi/k_z)\vdel\cdot\vBp$}. For each $k_z$, $\vBp$ will satisfy a
closed equation with $k_z$ appearing as a parameter and no mode coupling in
$k_z$.

We now seek the solutions of \Eqref{induction} again in the form of a
superposition of shearing waves,
\mbox{$\vBp=\sum_{\vk_0}\vBk(t)\,\rme^{\rmi\vk(t)\cdot\vec{r}}$}, where the
perpendicular wave numbers $\vk_0$ and $\vk(t)$ are defined in the same way as
in the velocity decomposition [\Eqref{shwaves_decomp}]. $\vBk$ satisfies
\begin{align}
\nonumber
\partial_t\vBk &= S\Bxk\ey - \eta\left[k^2(t)+k_z^2\right]\vBk\\
\nonumber 
&+ \sum_{\vk_0'}\Phikk[\ez\times\vk'(t)]
\cdot[\vk(t)\unity - \unity\vk'(t)]\cdot\vBkk\\
&- \rmi k_z\sum_{\vk_0'}\uzkk\vBkk,
\label{Bk_eq}
\end{align}
where $\unity$ is a unit dyadic. We take the large-scale mean field to be the
$xy$-average of the total magnetic field, i.e.\ $\vBo=\vBp_0$. The dynamical
equation for the mean field is given by the $\vk_0=0$ component of
\Eqref{Bk_eq}:
\begin{multline}
\label{Bo_eq}
\partial_t\vBo =
S\Bxo\ey - \eta k_z^2\vBo - \rmi k_z\sum_{\vk_0}\uzk\vBmk\\
- \sum_{\vk_0}\Phik\left[\ez\times\vk(t)\right]\vk(t)\cdot\vBmk
\end{multline}
(note that $\bar{B}_z=0$ because $\vdel\cdot\vBo=\rmi k_z\bar{B}_z=0$). 

We now calculate $\vBmk$ in \Eqref{Bo_eq} in terms of $\vBo$ via
\Eqref{Bk_eq}. This is particularly easy in the limit
$\Rm\ll\min(1,\St,\Sh^{-1})$, where $\Rm\sim\urms\lf/\eta$ and
$\Sh\sim{}S\lf/\urms$. We also assume $k_z\lf\ll1$, which will be verified
\emph{a posteriori} for the fastest growing dynamo mode. With these
approximations, the dominant terms in \Eqref{Bk_eq} are $\eta k^2(t)\vBk$ and
the $\vk_0'=\vk_0$ components of the wavenumber sums, giving 
\begin{equation}
\vBmk =
-\frac{\rmi k_z\uzk^*\vBo + \Phik^*\left[\ez\times\vk(t)\right]\vk(t)\cdot\vBo}
{\eta k^2(t)}.
\end{equation}
Substituting this into \Eqref{Bo_eq}, we get
\begin{equation}
\partial_t\vBo = S\Bxo\ey - [\eta + \beta(t)]k_z^2\vBo
+ \rmi k_z\ez\times\talpha(t)\cdot\vBo,
\label{alpha_eq}
\end{equation}
where $\beta(t)=\sum_{\vk_0}|\uzk(t)|^2/\eta k^2(t)\ll\eta$ (negligible
``turbulent diffusivity'' in the limit of low $\Rm$) and 
\begin{equation}
\label{alpha_expr}
\talpha(t) = 2\sum_{\vk_0}\frac{\uzk(t)\Phik^*(t)}
{\eta k^2(t)}\vk(t)\vk(t).
\end{equation}
\Eqref{alpha_eq} has the form of a standard mean-field equation
\cite{Steenbeck_Krause_Raedler,*Moffatt_book} with mean electromotive force
\mbox{$\vec{\mathcal{E}}=\talpha\cdot\vBo-\rmi
k_z\beta\,\vec{e}_z\times\vBo$}, but it is of stochastic nature: $\talpha(t)$
and $\beta(t)$ fluctuate with the correlation time $\tcorr$ of the velocity
field. Note that \mbox{$\langle\talpha(t)\rangle=0$} because we have
constructed our velocity field in such a way that
\mbox{$\langle\uzk(t)\Phik^*(t)\rangle=0$} (cf.\
\cite{RK08,Brandenburg_etal_stoch_alpha,
Vishniac_Brandenburg,Kraichnan1976,*Silantev,*Proctor}).

If we now average \Eqref{alpha_eq} over forcing realizations and look for
exponential growth of $\langle\vBo(t)\rangle$, we will find that, under the
approximations we have made, no such growth occurs to the lowest order in the
standard cumulant expansion \cite{vanKampen2} used to calculate
$\langle\talpha(t)\cdot\vBo(t)\rangle$
\footnote{With ${\langle}\talpha(t){\rangle}=0$, the cumulant expansion
\cite{vanKampen2} carried out to second order in $\Ku\sim k_z\lf\Rm\,\St$
yields ${\langle}\talpha(t)\cdot\vBo(t){\rangle}={\langle}\talpha(t)\cdot
\!\int_0^\infty\!\rmd t' \vec{\hat{G}}(t') \cdot[\rmi
k_z\ez\!\times\talpha(t-t')]{\rangle}
\cdot\vec{\hat{G}}(-t')\cdot{\langle}\vBo(t){\rangle}$, where $t\gg\tcorr$ and
$\vec{\hat{G}}(t)=\unity+St\vec{e}_y\vec{e}_x$. This vanishes on account of
\Eqsand{Phi12_corr}{uz12_corr} because
$\vec{k}(t)\cdot\vec{\hat{G}}(t')\cdot[\vec{e}_z\times\vec{k}(t-t')]=0$ for
any laboratory-frame wave vector $\vec{k}(t)$. See
also~\cite{SridharSubramanian2009a,*SridharSubramanian2009b,
*SridharSingh2010,*SridharSingh2011} for a different calculation leading to a
similar result.}.
While it is possible that the mean field grows at a higher order in the
expansion, the key question that needs to be addressed at lowest order is, in
fact, not necessarily whether the statistical average of the large-scale field
$\vBo(t)$ exhibits exponential growth, but whether the mean large-scale
magnetic energy $\langle|\vBo(t)|^2\rangle/2$ does. 

\paragraph{Large-Scale Energy.}
In order to address this last question, we introduce the mean-field covariance
vector
\begin{equation}
  \vC = (\Bxo^*\Bxo^{\vphantom{*}},
              \Byo^*\Byo^{\vphantom{*}},
              \Re\Bxo^*\Byo^{\vphantom{*}},
              \Im\Bxo^*\Byo^{\vphantom{*}}),
\end{equation}
where $\Re$ and $\Im$ denote real and imaginary parts. The evolution equation
for $\vC$ follows directly from \Eqref{alpha_eq}: 
\begin{equation}
  \label{covariance}
  (\partial_t + 2\eta k_z^2)\vC = (\tS + k_z\tA)\!\cdot\vC,
\end{equation}
where we have introduced the matrices
\begin{equation}
  \tS = \left[\begin{smallmatrix}
    0 & 0 & 0  & 0\vphantom{\alpha_{yy}} \\
    0 & 0 & 2S & 0\vphantom{\alpha_{yy}} \\
    S & 0 & 0  & 0\vphantom{\alpha_{yy}} \\
    0 & 0 & 0  & 0
  \end{smallmatrix}\right]
  \ \ \mathrm{and}\ \ 
  \tA = \left[\begin{smallmatrix}
    0 & 0 & 0 &  2 \alpha_{yy} \\
    0 & 0 & 0 &  2 \alpha_{xx}\vphantom{\alpha_{yy}} \\
    0 & 0 & 0 & -2 \alpha_{xy} \\
    \alpha_{xx} & \alpha_{yy} & 2\alpha_{xy} & 0
  \end{smallmatrix}\right].
\end{equation}
In the following,
we will use arabic numerals to refer to the components of the vectors and
matrices in \Eqref{covariance}.

We now average \Eqref{covariance} with respect to forcing realizations using
the cumulant expansion \cite{vanKampen2} to calculate
\mbox{$\langle\tA(t)\cdot\vC(t)\rangle$}. Since the Kubo number
\mbox{$\Ku\sim{}k_z\tA\tcorr\sim{}k_z\lf\Rm\,\St\ll1$}, the expansion can be
truncated at the lowest order in $\Ku$. The result is that
$\langle\vC(t)\rangle$ satisfies, for $t\gg\tcorr$, 
\begin{equation}
  (\partial_t + 2\eta k_z^2)\langle\vC\rangle =
  (\tS + k_z^2\tD)\cdot\langle\vC\rangle,
\end{equation}
where the term originating from $\talpha$ now has the form 
of a (negative) tensor diffusivity
\begin{align}
  \label{Dmatrix}
  \tD = \int_0^\infty\!\!\rmd t'\langle\tA(t)\cdot\tA(t-t')\rangle.
\end{align}
We have also assumed \mbox{$S\tcorr\ll1$}, which allowed us to neglect the
matrix exponentials of $\tS t'$ in \Eqref{Dmatrix}.

We note that $\tD$ is block diagonal: its elements are zero where those of
$\tA$ are not, and vice versa. It follows that \mbox{$\langle
C_4\rangle=\langle\Im\Bxo^*\Byo^{\vphantom{*}}\rangle$} evolves independently
of the other components of $\langle\vC\rangle$: 
\begin{equation}
  \partial_t\langle C_4\rangle = -k_z^2(2\eta - D_{44})\langle C_4\rangle.
\end{equation}
Since $D_{44}/\eta\sim\Rm^3\St\ll1$
\footnote{In fact, it is not hard to
prove, using \Eqref{alpha_expr}, then \Eqsand{Phi12_corr}{uz12_corr}, that
$D_{44}=0$ exactly.},
we conclude that $C_4$ always decays, which means that $\Bxo$ and $\Byo$
asymptotically have the same complex phase.  

With $C_4=0$, we are left with a rank-three eigenvalue problem. If we let
$\langle\vC\rangle\propto \exp(2\gamma t)$, the resulting dispersion relation
will be a cubic equation in $\gamma+\eta k_z^2$. This equation can be solved
perturbatively in the limit $k_z^2\tD \sim k_z^2\tA^2\tcorr \ll S$ or,
equivalently, \mbox{$(k_z\lf)^2\, \Rm^2\,\St\,\Sh^{-1}\ll1 $}. In the end,
this means that the only element of the tensor $\tD$ that survives to give a
non-negligible contribution is 
\begin{equation}
  \label{D12}
  D_{12} = 2\int_0^\infty\!\!\rmd t'\bigl\langle\alpha_{yy}(t)\alpha_{yy}(t-t')
  \bigl\rangle,
\end{equation}
where $t\gg\tcorr$. Then $\gamma$ satisfies 
\begin{equation}
  \label{leading_balance}
  (\gamma + \eta k_z^2)^3 - \frac{k_z^2 S^2 D_{12}}{4} = 0
\end{equation}
and so, assuming that $D_{12}>0$, the real root of this equation gives the
dynamo growth rate \footnote{The non-resistive part of the dynamo growth rate
\Exref{growth_rate} can also be obtained via heuristic arguments given in
\cite{Vishniac_Brandenburg}, see their Eq.~(13).}
\begin{equation}
  \label{growth_rate}
  \gamma = -\eta k_z^2 + \left(\frac{k_z^2 S^2 D_{12}}{4}\right)^{1/3}.
\end{equation}
The vertical wave number and the growth rate of the fastest growing mode are
\begin{equation}
  \label{kpeak}
  \kpeak = \frac{|S|^{1/2}}{\sqrt{2}}
  \left(\frac{D_{12}}{27\eta^3}\right)^{\!1/4}
  \!\!\!\sim \lf^{-1}\Sh^{1/2}\,\St^{1/4}\,\Rm^{5/4}
\end{equation}
(confirming $\kpeak\lf\ll1$) and 
\begin{equation}
  \label{gmax}
  \gamma_\mathrm{max} =
  \frac{|S|}{3}\!\left(\frac{D_{12}}{3\eta}\right)^{\!1/2}
  \!\!\sim \frac{\urms}{\lf}\,\Sh\,\St^{1/2}\,\Rm^{3/2}.
\end{equation}
The structure of this mode is such that
\begin{equation}
  \frac{\langle|\Bxo|^2\rangle}{\langle|\Byo|^2\rangle} = \frac{D_{12}}{6\eta}
  \sim\St\,\Rm^3,
\end{equation}
which is independent of shear. 

\paragraph{Discussion.}
The scalings derived above, viz., \mbox{$\kpeak\propto S^{1/2}$},
\mbox{$\gamma_\mathrm{max}\propto S$}, and the independence of
\mbox{$\langle|\Bxo|^2\rangle/\langle|\Byo|^2\rangle$} of $S$, are precisely
the ones reported in the numerical experiments
\cite{Yousef_etal_PRL,Yousef_etal_AN,Kapyla_etal,*Hughes_Proctor}. One should
keep in mind that most of these simulations were not done in the asymptotic
regime \mbox{$\Rm\ll\min\left(1,\St,\Sh^{-1}\right)$} or at particularly small
$S\tcorr$. The fact that the scalings we have derived nevertheless appear to
hold even for parameter values at the boundary of the analytically tractable
regime might be interpreted as a testimony to the robustness of the underlying
physical effect
\footnote{Note that while the distinction between the growth of the
large-scale magnetic \emph{vector} field and of the large-scale magnetic
\emph{energy} was not appreciated by \cite{Yousef_etal_PRL,Yousef_etal_AN},
their numerical results on the dependence of the field's growth rate and
vertical scale on $S$ all referred to the root-mean-square large-scale field
integrated over the numerical box and so a comparison with our predictions for
the large-scale energy is appropriate. As the vector large-scale field at any
given $z$ randomly changed sign in their simulations on timescales of order a
few growth times, it is plausible that its long-term time average was indeed
zero, as suggested by our theory, although this was not checked at the time.
A more extensive investigation of the feasibility 
of mean-field growth in broader parameter regimes than attempted here 
can be found in \cite{McWilliams}}.
Indeed, we note that  \Eqref{growth_rate} rigorously holds for any
``quasi-2D'' velocity field superimposed on a uniform shear flow with the only
provisos that it has a well defined characteristic length scale, a correlation
time much shorter than the inverse rate of shear, and the property that
\mbox{$\langle\uzk\Phik^*\rangle=0$}. Our theory shows that such a velocity
field is always capable of dynamo action provided sufficiently large scales in
the $z$-direction are accessible to the mean field (i.e., provided the system
is large enough). A field of randomly forced shearing waves at low $\Rey$,
given by \Eqsand{Phi_eq}{uz_eq}, is a physically realizable example of such a
velocity field. For this field, using \Eqsref{Phi12_corr}, \Exref{uz12_corr},
\Exref{alpha_expr}, \Exref{D12}, and $S\tcorr\ll1$, we get
\begin{equation} 
\label{D12_expr}
D_{12} = 4\sum_{\vk_0} k_y^4 
\frac{\bigl<|\Phik(t)|^2\bigr>\bigl<|\uzk(t)|^2\bigr>}{\nu\eta^2k^6(t)},
\end{equation} 
which is positive, as assumed in \Eqref{growth_rate}. 

The key ingredient in the dynamo loop are fluctuations in the $\talpha$-tensor
\Exref{alpha_expr}, which, in conjunction with stretching of the mean field by
the background shear flow, provide a positive feedback \footnote{Comparing
\Eqsand{helicity}{alpha_expr}, we see that fluctuations in $\talpha$ are
related to fluctuations in kinetic helicity, although, for a general multiscale
velocity field, the correspondence is by no means exact (cf.\
\cite{CattaneoHughes2006}).}. This is evocative of the dynamo models known as
the ``stochastic $\alpha$-effect'', which are based on introducing a
fluctuating scalar $\alpha_{yy}$ \cite{RK08,
Vishniac_Brandenburg,Kraichnan1976,*Silantev,*Proctor} --- this has usually
been done based on \emph{ad hoc} non-rigorous models of how this $\alpha$
comes about. The theory we have presented here is the first calculation of
this kind done from first principles.

\paragraph{Conclusion.}
We have presented a minimal analytically tractable model of the shear dynamo.
The simplicity of the model suggests that the effect is robust, while its
rigorous validity in the realizable limit of low $\Rm$, weak shear and for a
velocity field consisting of randomly forced shearing waves at low $\Rey$
suggests that it is physical and does not depend on \emph{ad hoc} closure
assumptions. Much remains to be understood before it can be assessed whether
the shear dynamo offers a panacea for (non-helical) generation of large-scale
magnetic fields in astrophysical systems. A further effort in this direction
appears worthwhile in view of the great success enjoyed by shear-induced
dynamos in astrophysically motivated numerical experiments and a basic
similarity of the field structure that they generate
\cite{Rincon_Ogilvie_Proctor,*Gressel_etal,*Lesur_Ogilvie,
Yousef_etal_PRL,Yousef_etal_AN,Kapyla_etal,*Hughes_Proctor}. A companion paper
on the quasilinear elemental shear dynamo, exploring broader parameter
regimes, is \cite{McWilliams}. A major outstanding task is to understand how
the shear dynamo mechanism of generating large-scale fields coexists with the
fluctuation dynamo of small-scale fields, which will inevitably be present at
sufficiently large $\Rm$ \cite{Yousef_etal_AN} and, therefore, in any real
astrophysical situation (cf.\ \cite{Boldyrev_Cattaneo_Rosner}). 

\begin{acknowledgments} 
This work was supported in part by NSF grant AST--0807432 (T.H.), NASA grant
NNX08AH24G (T.H.), STFC (A.A.S.) and by the Leverhulme Trust Network for
Magnetized Plasma Turbulence. Important discussions with A.~Brandenburg,
N.~Kleeorin, G.~Lesur, A.~Mallet, I.~Rogachevskii, and T.~Yousef at various
stages of this work's gestation are gratefully acknowledged. 
\end{acknowledgments}

\bibliography{shear-dynamo}

\end{document}